\documentstyle[worldsci,12pt]{article}
\def\lsim{{\mathop <\limits_\sim}}

\begin{document}
\begin{titlepage}
\begin{flushright}
UM-TH-93-10\\
hep-ph/9304244\\
April 12, 1993
\end{flushright}
\vspace*{1.7in}
\begin{center}
{\Large
Is the World Supersymmetric?  Do We Already Know?\footnote{Based on
Invited Talks at the Coral Gables Conference on ``Unified Symmetry in
the Small and in the Large'', Jan. 1993, and at the XVIII Rencontres
de Moriond, Les Arc, March, 1993.}}
\vspace{1cm}\\
{\large G.L. Kane}
\vspace{1cm}\\
{\em Randall Physics Laboratory}\\
{\em University of Michigan}\\
{\em Ann Arbor, MI  48109-1120, USA}
\vspace{.5cm}\\
{\large Abstract}
\end{center}
\begin{quotation}
In addition to the very good theoretical motivations for
supersymmetry, there are now at least nine phenomenological
indications that nature is supersymmetric.  All are indirect, so more
is better.  They are enumerated here.  Some discussion is also given
of models, of when and where superpartners might be directly detected,
and of why the scale of supersymmetry cannot be pushed up if
superpartners and SUSY Higgs bosons are not directly detected.
\end{quotation}

\end{titlepage}
\newpage
\noindent {\bf INTRODUCTION}
\smallskip

It is widely accepted that new physics, beyond the Standard Model
(SM), will be discovered even though the SM successfully describes
experiments.  Reasons for believing there will be new physics fall
into at least four categories:\ \ regularities not explained by the
SM, such as $Q_p = - Q_e$, the similarity of quark and lepton spectra,
mass relations such as $m_b = m_\tau$ at the GUT scale, the
unification of gauge couplings at the GUT scale, etc.; dynamical
questions such as the origin of mass and the hierarchy of masses;
``why'' questions such as why three families, why $SU(3)\times
SU(2)\times U(1),$ why is parity violated; and connections to gravity
and the cosmological constant problem.  Some might include as an
additional category the tantalizing properties of some ambitious
theories that include possible answers to some of the above questions,
or promising approaches to them.

Given that new physics will occur, one might wonder if we already have
clues to what the new physics is.  Can we work it out without needing
to see it explicitly?  Physicists are supposed to be good at indirect
reasoning.

Of the alternatives that have been considered probably the largest
number of theorists expect nature to be supersymmetric, but
theoretical prejudices do not have a great record of being right.  The
theoretical motivations for supersymmetry, particularly the connection
to gravity and the ability to maintain two widely separated scales,
are powerful but do not guarantee that nature is supersymmetric.

I believe that in the past two years indirect evidence for
supersymmetry has been accumulating.  There are now at least nine
\underbar{phenomenological} indications, i.e., phenomena that are
predicted by or consistent with supersymmetry but would not have been
if some data had been different.  Five are things that happen and
four are things that do not happen.

Since these constitute indirect evidence, more is better.  Obviously
any single indirect piece of evidence cannot uniquely point toward
supersymmetry.  Once superpartners are found it will not require
special cleverness to accept supersymmetry.  It is more fun to get it
right ahead of the direct evidence.  So the familiar (Feynman?) remark
that one good argument is better than several bad ones not only does
not apply here, it is misleading.  Indeed, anyone who would prefer
``one good argument'' should pause and ask whether there is any
indirect argument that would convince them.

A number of groups of people have been studying models that
increasingly build in theoretical and phenomenological constraints.  I
will not review these here, but it is relevant to know that one can
construct real models that exhibit all of the properties described
below.  Groups studying models include R. Arnowitt and P. Nath; R.
Roberts and G.G. Ross; J. Lopez, D. Nanopoulos, A. Zichichi, H. Pois
et al.; J. Hagelin, S. Kelley et al.; J. Ellis, F. Zwirner et al.;
G.L. Kane, Chris Kolda, L. Roszkowski, James Wells; R. Barbieri et al.; P.
Langacker et al.; Q. Shafi et al.; T. Yanagida, H. Murayama et al.; M.
Carena, C. Wagner, S. Pokorski.

The spirit in which this talk is offered is to encourage more people
to take supersymmetry more seriously.  Whenever a large number of good
physicists have taken a topic seriously and focussed on it, progress
has been rapid.  There is much need for theoretical work to improve
the formulation of the theory at the electroweak scale, and to study
how supersymmetry is broken.  Experimenters could increase priorities
for direct and indirect searches motivated by supersymmetry
predictions, and pay particular attention to sensitivity to
supersymmetry triggers and signatures.

None of the following indications are unique --- each could have an
alternative interpretation.  Taken together, they relate a large
amount of beyond-the-SM physics.  No other approach comes close to
such achievements.  In fact, other approaches may be most valuable to
show that it is easy to get wrong the things that supersymmetry gets
right.  Perhaps a conservative view is that the items below are all
tests of any approach to physics beyond the SM.

All of the indications discussed below have also been remarked on by
others.  This is a review talk, bringing these points together.  It is
also important to be sure that all of these effects can occur in one
consistent model.  The whole is greater than the sum of the parts.

\medskip
{\noindent\bf THINGS THAT HAPPEN}
\medskip

\noindent A.\ \  It is now familiar \cite{langacker,amaldi} to everyone that
as the precision of measurement of the gauge couplings at $m_Z$
improved, the gauge couplings still could be unified at a scale we can
call the GUT scale, about $10^{16}$ GeV, in a supersymmetric theory.
Although there is another parameter in a supersymmetric theory (the
scale of superpartner masses), it is not arbitrary --- unless it is
between about 0.1 TeV and 1 TeV it is not the supersymmetric theory of
interest. It comes out right.  Unification of the three gauge
couplings does not occur\cite{langacker,amaldi} at all in a non-SUSY
$SU(5)$ GUT.  It occurs in some other theories, but only if an
intermediate scale is added.  Our \cite{kolda} version of the curves
is shown in Fig. 1.

\medskip

\nocite{arason}
\noindent B.\ \ It was pointed out in ref. 4 that independently the coming
together of $m_b$ and $m_\tau$ at the GUT scale occurred in the SUSY case
and not in the non-SUSY case.  Our version of this is also shown in
Fig. 1.  What is significant is that $m_b = m_\tau$ at the same scale
where the gauge couplings meet, within experimental errors.  Given the
apparent unification of forces, and the apparent similarity of quark
and lepton spectra, it would be courageous to claim this result  is an
accident.

\bigskip

\noindent C.\ \   It was noticed \cite{review} a decade ago that {\bf if}
$m_{\rm top}$ were large enough (somewhat above $m_W$), then the Higgs
mechanism occurs automatically in a supersymmetric theory, rather than
having to be an ad hoc add-on as in the SM.  $m_{\rm top}$ is indeed
that large.  This is a beautiful result.  The coefficient $m_{h_t}^2$
of the quadratic term in the scalar potential for the scalar field
coupled to up-type quarks goes negative and triggers the Higgs
mechanism.  The weak scale is determined by where $m_{h_t}^2$ goes
negative, and in the same models that give Fig. 1 (and other
quantitative results described below) this occurs in such a way that
$m_Z^2$ can be fitted accurately.  While parameters are adjusted
to get $m_Z^2$ right, there is no fine tuning if the SUSY mass
parameters are well below 1 TeV.

Fig. 2 shows how this happens, and shows the SUSY spectrum in one
model.  At the GUT scale the masses have a simple pattern.  At the
weak scale they are calculated to have other values.  One (mass)$^2$
of a scalar goes negative, the one whose renormalization group
equation contains a large negative term from the top quark Yukawa
coupling (imagine the (mass)$^2$ as running from the GUT scale down).

\bigskip

\noindent D.\ \  While discovery of a light Higgs boson would not convince
everyone that nature is supersymmetric, it would satisfy most.  Global
analyses \cite{ellis} of the precision measurements from LEP
and FNAL have typically shown a mild preference for a light Higgs
boson when analyzed as functions of $m_{\rm top}$ and $m_h$.  For our
purposes ``light'' could be any $m_h < 1$ TeV; once a Higgs boson
exists, in a world with a high scale where the theory remains
perturbative between the high scale and the weak scale,
theorems~\cite{cabibbo,kane,espinosa} require $m_h \lsim 170$ GeV in
the SM and $m_h \lsim 146$ GeV for SUSY, for the relevant region of
$m_{\rm top}$; A and B above show that the existence of the high scale
is now no longer a matter of taste.

The relevant analysis of the data to decide if the precision
measurements provide evidence for a light Higgs boson is not the two
variable one, but one where $m_{\rm top}$ is fixed.  Let us assume the
candidate events being observed at FNAL will indeed eventually lead to
reporting the detection of $m_{\rm top}$.  Then from the integrated
luminosity taken by the detectors, and the published cross section,
one would guess $m_{\rm top} \cong 135$ GeV, a value fully consistent
with the LEP data (remember, if one is testing the hypothesis of a
light Higgs one should use the LEP data analysis with $m_h \cong
m_Z$).  Once $m_{\rm top}$ is fixed at such a value, one
finds\cite{delaguila}that the data prefers a ``light'' Higgs boson at
the level of several standard deviations.

When analysis of the data now in hand at LEP is completed this result
should become firmer if we are not being fooled by statistical or
unknown systematic effects.  And soon the detection (and mass) of top
should be settled if indeed $m_{\rm top} \sim$ 135 GeV.  Until then,
we can tentatively conclude that there is mild evidence for the
existence of a Higgs boson, which can be interpreted as evidence for
supersymmetry.

\bigskip

\noindent E.\ \ The final thing that ``happens'' is the dark matter.
It has been understood for a decade that the LSP will normally be
stable, and will automatically be a good candidate for cold dark
matter.  As constraints from theory and data have increasingly been
imposed this has continued to be true.\cite{leszek} It is
phenomenologically non-trivial; for example, the lightest sneutrino is
excluded as a dark matter candidate but the lightest neutralino
behaves just right, giving $\Omega \sim 1$; the LSP with the right
properties is mainly gaugino.

One might ask ``what about R-parity violation that could make the
LSP unstable''?  Eventually phenomenological constraints may be able to
show that R-parity conservation is valid, or sufficiently valid to
give LSP's lifetimes larger than the age of the universe.  In some
theories \cite{ibanez,martin} R-parity conservation is natural.  But I
think the right answer to this question is that supersymmetry
automatically provided a dark matter candidate with certain properties
and this turned out to be the kind of dark matter needed by cosmology
and astrophysics.  In the context of the present analysis it is
appropriate to presume this is no accident; thus R-parity is
conserved.  In supersymmetric grand unified theories in general $\nu$
masses will exist, and perhaps also pseudo-Goldstone bosons from
breaking of global symmetries, so that other forms of dark matter will
also exist.  As we learn more about the theory, the ratios of the
contributions to $\Omega$ of the different forms will be calculated.

\medskip
\noindent{\bf THINGS THAT DON'T HAPPEN}
\medskip

\noindent F.\ \ Given the unification of forces, and $m_b=m_\tau,$ we
expect some kind of quark and lepton unification, and it is likely
that protons will be unstable.  Then it is well established that
proton decay is too rapid in some theories including an $SU(5)$ GUT, but
in a supersymmetric GUT the proton lifetime is OK.\cite{hisano} This
happens in part because the extra particle content of the
supersymmetric theory moves the GUT scale higher and the lifetime
scales as $m_{\rm GUT}^4$, but it is more complicated because the
dominant operators change.  Typically models give a range of
parameters for which the lifetime is OK, and a range for which it is
not, a nice situation because it suggests that experiments may be on
the edge of detecting a result.

\bigskip

\noindent G.\ \ It is well known that the precision measurements from
LEP and FNAL are well described by the SM if $m_{\rm top}$ and $m_h$
are in certain ranges.  People have for several years\cite{lynn} looked
for the effects of physics beyond the SM, in hopes that a clue would
appear or that some ideas would be constrained.  It is also well known
that some models indeed can give predictions in conflict with the data
--- most simply, a degenerate quark doublet gives a contribution
$+1/2\pi$ to the observable $S$, whose central value is somewhat
negative; and perhaps TC theories are inconsistent with the
data.\cite{peskin}  My point is not to argue one way or the other
about TC, but to point out that in the case of supersymmetry the
agreement with the precision measurements is, if anything,
better\cite{barbieri} than for the SM.  For much of the parameter
space the region predicted by SUSY and the $2\sigma$ experimental
region lie on top of each other.  The point of mentioning the other
models is simply to show that this is a non-trivial success.

\bigskip

\noindent H.\ \ It has been known\cite{hagelin} for a decade that
supersymmetric theories could give large flavor changing neutral
currents (FCNC) and induce $K^0 - \overline{K}^0$ mixing, $B^0 -
\overline{B}^0$ mixing, $K_L\rightarrow \mu e$, etc.  Such effects can
be a major problem for other kinds of beyond-the-SM-physics.  What
happens in supersymmetric models is that any model where the down-type
squarks are nearly degenerate at the GUT scale, and their splitting at
the weak scale arises from running their masses to the weak scale,
automatically satisfies FCNC constraints.  This is ``natural'', in the
sense that once the sfermions are approximately degenerate at the high
scale, this degeneracy is preserved by radiative corrections.  Thus
the smallness of FCNC can be viewed as a success of SUSY.

\bigskip

\noindent I.\cite{newhaber}\ \ In a supersymmetric theory the mass of
the lightest Higgs boson is calculable if the soft-breaking parameters
are known, and it depends on the same parameters as the superpartners.
In constrained models it is never very light.  In addition, the
one-top-loop radiative corrections\cite{morenewhaber}shift
$m_{h^\circ}$ up by $\sim 10$ GeV.  In practice $m_{h^\circ}$ is
usually in the range 70-110 GeV in models.  So if a light $h^\circ$
had been found at LEP already it would have been very difficult to
make it consistent with a supersymmetric world in which the other good
features listed in this talk were maintained simultaneously.

\newpage
\noindent{\bf WHEN, WHERE MIGHT SUPERPARTNERS OR SUSY HIGGS BE DETECTED?}
\medskip

This question is best studied with models, and such studies are in
progress by the groups mentioned in the introduction.  It is a
complicated and subtle task to incorporate all of the theoretical
and experimental constraints into models.  Progress is being made,
though the results of all groups are not yet consistent.  It is
remarkable that fully detailed models can indeed be constructed.  In
the next few months studies of models are likely to converge to a
fairly small region of allowed mass ranges.

At LEP there is a window for the lightest SUSY Higgs boson.  Typically
in models it ranges from 70 GeV to 110 GeV.  There is an upper limit
of about 146 GeV (see next section), but the upper limit is seldom
saturated in models.  LEP also has a window for the lightest chargino.
Dark matter constraints suggest\cite{leszek} sleptons may be
beyond the reach of LEP200.

At FNAL the decay $t\rightarrow bH^+$ can occur for a small region of
parameter space.  For higher luminosities production of gaugino pairs
is detectable in lepton channels.\cite{frere}  Over some of parameter
space the squarks and gluinos can be detected.  Particularly at larger
luminosities there is a significant window.

SSC and LHC will be supersymmetry factories that will permit the
detection and study of much of the spectrum, given appropriate
detectors.  Careful study of neutralinos and charginos, and detection
of sleptons and sneutrinos, may require NLC, which can do very useful
analyses, particularly with a polarized beam.

Models show clearly that superpartner masses fall naturally in regions
such that we would have been very lucky if a superpartner or SUSY
Higgs boson had been detected so far.  For LEP200, or FNAL with the main
injector, a little luck is still required.  Detailed results will be
reported in ref. 3, and have already been given for some cases by
several groups of authors listed in the introduction.  Although
sometimes people have mentioned the absence of superpartners as an
argument against supersymmetry, it is clear that constrained model
studies require superpartners typically heavier than $m_W.$

\bigskip
\noindent {\bf CAN THE MASSES OF SUPERPARTNERS AND SUSY HIGGS BE\\
PUSHED UP IF THEY ARE NOT DETECTED?}
\medskip

What if superpartners or a light Higgs boson are not detected?  Can
supersymmetry be excluded, or will it just be pushed to a higher
scale?  The answer is that it can be tested definitively.  There are
three relevant points; (i) refers to the lightest Higgs boson and
(ii) and (iii) to superpartners.

\begin{description}
\item{(i)} It has been possible to show\cite{kane,espinosa} that
there exists an upper limit on the mass of the lightest Higgs boson in
any supersymmetric theory.  [``Any'' could include extensions of the
Higgs sector, the low energy gauge group, and the spectrum of heavy
fermions.  The limit exists including all of these.]  Its value has
been calculated\cite{kane,espinosa} (within a few \% accuracy) if the
low energy gauge group is the SM one; it is 146 GeV for $m_{\rm top} >
100$ GeV.  [Additional heavy fermions could increase this \cite{moroi}
as much as 12 GeV per family; other constraints suggest at most two
such families could exist.  Most likely no additional fermions exist.
If the gauge group is larger,\cite{haber} I think the upper bound
turns out to be lower, but that is not proved yet.]  Thus either a
light Higgs boson is detected below this bound, or all of our
indications are accidents and supersymmetry is unconnected to
understanding physics at the weak scale.  In models the bound is
seldom saturated; typically $70 \lsim m_{h^\circ} \lsim 110$ GeV.

\item{(ii)} There are two kinds of arguments that superpartner masses
cannot be pushed up.  Today both can be barely evaded by unpleasant
fine-tunings, but it seems likely that continued study of models and
constraints will sharpen this situation.

\item{\ \ \ } One argument is that unless some slepton or squark is
light enough (a slepton in models) there will be too much dark matter,
and the universe will be overclosed.  The LSP's are in equilibrium
with other particles in the early universe.  They can annihilate by
sfermion exchange $(\tilde f)$ to final fermions $({\rm LSP} +
\overline{{\rm LSP}} \rightarrow f + \overline{f})$, or via an
s-channel $Z, {\rm LSP} + \overline{{\rm LSP}} \rightarrow Z^{(*)}
\rightarrow f + \overline{f}$; other final states are typically less
important.  Since the $Z$ couplings are neutral current ones, giving
small rates, except for special choices of the LSP wave function
(mainly Higgsino LSP's since $\gamma$ and $B$ do not couple to $Z$,
and such LSP's are disfavored\cite{leszek}) the annihilation through
$Z$ is too small to avoid giving $\Omega >1.$ If sfermions are too
heavy, the annihilation rate by sfermion exchange will be too small to
bring $\Omega$ to unity.  Numerically,\cite{leszek} sleptons below
about 400 GeV are needed.  Since most superpartner masses are tied
together, this implies most of the superpartners will be light enough
for SSC/LHC to detect.  Note that this argument does not require a
committment to how much dark matter is due to the LSP, but only that
the dark matter does not overclose the universe.

\item{(iii)} The second kind of argument is based on models.
Basically, satisfying the theoretical and experimental constraints
always produces a spectrum that is at least in part detectable at
SSC/LHC.  A recent analysis\cite{roberts} found upper limits on all
superpartner masses without requiring fine tuning constraints; all
masses were well within the SSC/LHC range.  It is possible that in the
near future combining a number of constraints will allow a firm upper
limit to be set on superpartner spectra from acceptable supersymmetric
theories.\cite{kolda}

\end{description}
\vspace*{1cm}
\noindent{\bf ACKNOWLEDGEMENTS}
\medskip

I appreciate comments and assistance from C. Kolda, J. Wells, L.
Roszkowski, H. Haber, M. Dine and D. Kennedy.

\newpage

\begin{center}
{\Large Figure Captions}
\end{center}
\bigskip

\noindent {\bf Figure 1} shows the running of the gauge couplings$^{(1,4)}$
$(\alpha_i^{-1})$, and independently of the masses$^{(4)}$ $m_b$ and
$m_\tau$, in theories with supersymmetric RGE's.  The left scale is
for $m_b, m_\tau$ and the right scale for $(\alpha_i^{-1})$.  The band
for $m_b$ is determined by the uncertainty in extracting $m_b$ from
data.\cite{arason} Note that both come together at the same high scale.

\bigskip

\noindent {\bf Figure 2} shows the spectrum of one model where there
is a simple spectrum at the GUT scale, and different superpartner
masses are generated at the weak scale by radiative effects.  Colored
superpartners automatically get heavier so color symmetry cannot be
broken.  One Higgs field (mass)$^2$ goes negative and produces the
Higgs mechanism as a consequence of the structure of the
supersymmetric theory.  For that field what is shown is
$-|m_{h_t}^2|^{1\over 2}$.


\begin{thebibliography}{99}
\bibitem{hagelin}
A new study of these effects, with references to the literature,
arrived as this was being written: John S. Hagelin, S. Kelley, and
Toshiaki Tanaka, MIU-THP-92/59.
\end{thebibliography}
\end{document}